\documentstyle[aps,multicol,epsfig,graphicx,subfigure]{revtex}
\title{Micellization in the presence of polyelectrolyte}
\author{N. Lemke}
\address{Programa Interdisciplinar de Computa\c c\~ao Aplicada \\
Centro de Ci\^encias Exatas e da Terra -- Unisinos \\
Av. Unisinos, 950 
93022-000 -- S\~ao Leopoldo -- RS, Brazil \\
E-mail: {\tt lemke@exatas.unisinos.br}}
\author{Jeferson J. Arenzon}
\address{Instituto de F{\'\i}sica, Universidade Federal
do Rio Grande do Sul\\ Caixa Postal 15051, 91501-970, Porto Alegre, RS,
Brazil  \\ E-mail: {\tt arenzon@if.ufrgs.br}}    
\author{Yan Levin\footnote{Corresponding author}}
\address{Instituto de F{\'\i}sica, Universidade Federal
do Rio Grande do Sul\\ Caixa Postal 15051, 91501-970, Porto Alegre, RS,
Brazil \\ E-mail: {\tt levin@if.ufrgs.br}}

\begin{document}
\maketitle

\begin{abstract}
We present a simple model to study micellization of amphiphiles
condensed on a rodlike polyion.  Although the mean field theory
leads to a first order micellization transition for
sufficiently strong hydrophobic interactions, the simulations
show that no such thermodynamic phase transition exists. 
Instead, the correlations between the condensed amphiphiles
can result in a structure formation very similar to micelles.

\end{abstract}

\pacs{
  \Pacs{05}{70.Ce}{Thermodynamic functions and equations of state}
  \Pacs{61}{20.Qg}{Structure of associated liquids: electrolytes, molten
                   salts,etc}
  \Pacs{61}{25.Hq}{Macromolecular and polymer solutions; polymer melts;
      swelling }
}

\section{Introduction}

Interaction between polyelectrolytes and ionic amphiphiles
has attracted significant attention in both the 
molecular biology and the condensed matter physics communities~\cite{fri97,hop98,fel97,fel89,fel91,ver97,saf97,har98,and98,shi87,gor98,dia00,dia00b}.  
The driving motivation for this surge in interest is the
possible application of polyelectrolyte-amphiphile complexation
in gene therapy~\cite{hop98}.  One of the major stumbling blocks
in designing a successful gene therapy is the lack of a transfection
mechanism by which a DNA strand  can be inserted into a 
cell~\cite{hop98}.  The problem arises as a result of strong electrostatic
repulsion between the DNA and the molecular membrane, both
of which are negatively charged.  This repulsion prevents
a DNA segment from coming in contact with a cell membrane,
thus precluding any possibility of transfection.
To overcome the electrostatic repulsion a number of 
protocols have been
developed.  The most explored ones rely on genetically
modified viral
vectors. A number of complications
which can arise from the viral gene therapy have stimulated a
development of non-viral methods~\cite{hop98}.  One such method explores
the association between a negatively charged DNA and cationic
lipids or surfactants.  A particular 
method which has attracted much
attention relies on formation of lipoplexes~\cite{fel97,fel89,fel91}.  
These are complexes
composed of a DNA strand and cationic lipid vesicles.  
Unfortunately the non-viral methods are also
prone to problems since, as is well known, the  
cationic surfactants and lipids are toxic to
an organism~\cite{hop98}. An interesting question which then arises is what
is the minimum concentration of cationic
amphiphile  needed to form a lipoplex or a surfoplex? This question
is well posed, since it has been known for some time that
the polyelectrolyte-ionic amphiphile complexation occurs in a 
cooperative manner~\cite{shi87,gor98,kuh98,kuh99,kuh99a}.  
It is, therefore, possible to identify 
the location of the cooperative condensation with the critical
concentration needed to form complexes.  What is the internal 
structure of such complexes?
Are the condensed amphiphiles uniformly distributed along the
DNA or do they form micellar aggregates on the surface of a polyion?
As a first attempt to study this difficult problem we shall appeal to a 
very simple model~\cite{kuh99b,kuh00}.

\section{The Model}


We consider a rigid polyion modeled as a cylinder of length $Zb$ 
and radius  $Rb$ inside a uniform medium  of dielectric constant $\epsilon$. 
The total  charge of a polyion, $-Zq$, is uniformly 
distributed along the length of the cylinder so that each one of the
$Z$ monomers has charge $-q$. 
To further simplify the calculations, the longitudinal
and angular degrees of freedom are discretized. The surface
of the cylinder is subdivided into $Z$ parallel rings of $n$ sites each.
The $m$ condensed amphiphiles are restricted to move between
the $Zn$ ring sites, on the surface of the cylinder.  

Each amphiphile has a charged head group and a hydrocarbon tail. 
For generality
we shall take the head group to have  charge  $\alpha q$.
The hardcore repulsion between the amphiphiles requires
that each site is occupied by at most one amphiphile.  
For the specific case of DNA and dodecyltrimethyl amonium bromide
(DoTAB) at the cooperative binding transition 
$m \approx 0.8Z$~\cite{gor98,kuh98}.

We define the occupation variables $\sigma_{ij}$,
with $i=1, \ldots, Z$ and $j=1,\ldots,n$, in such a way that $\sigma_{ij}=1$
 if a surfactant is attached at $j$'th ring position of the $i$'th
monomer and $\sigma_{ij}=0$ otherwise. Since the number of amphiphiles 
is fixed, the values of occupation variables obey
the constraint 
$$
\sum_{i=1}^Z \sum_{j=1}^n \sigma_{ij} = m\; .
$$

The Hamiltonian for this model has the following contributions:
\begin{itemize}
\item Electrostatic surfactant-monomers interaction:
\begin{equation}
\beta {\cal H}_1=-\alpha \xi\sum_{i,j=1}^{Z} \sum_{k=1}^{n}
\frac{\sigma_{ik}}{\sqrt{R^2+(i-j)^2}},
\label{e1}
\end{equation}
\end{itemize}
where we have introduced a dimensionless charge 
density,  the Manning parameter~\cite{man69,lev96,bar96},
\begin{equation}
\xi \equiv  \frac{\beta q^2}{b\epsilon} \;\;. 
\end{equation}

\begin{figure}[h]
\centerline{\epsfig{file=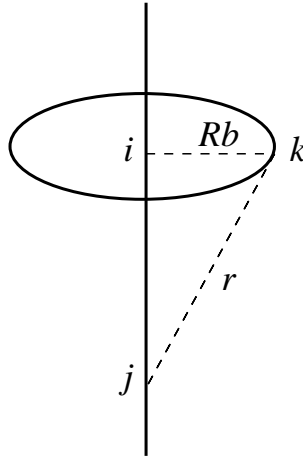,width=4cm,angle=0} }
\caption{Interaction between a surfactant in the $i$'th ring with 
the $j$'th monomer.  The distance between the surfactant and the
monomer is $r=b\sqrt{R^2+(i-j)^2}$.}
\label{figure1}
\end{figure}

\begin{itemize}
\item Electrostatic surfactant-surfactant interaction:
\begin{equation}
\beta{\cal H}_2=\frac{\alpha^2 \xi}{2}\sum_{i,j=1}^{Z} \sum_{k,l=1}^{n}
\frac{\sigma_{ik}\sigma_{jl}(1-\delta_{ij}\delta_{kl})}{
\sqrt{(i-j)^2+4R^2\sin^2\left(\frac{\pi}{n}
|k-l| \right)}} \;.
\label{e2}
\end{equation}

\begin{figure}[h]
\centerline{\epsfig{file=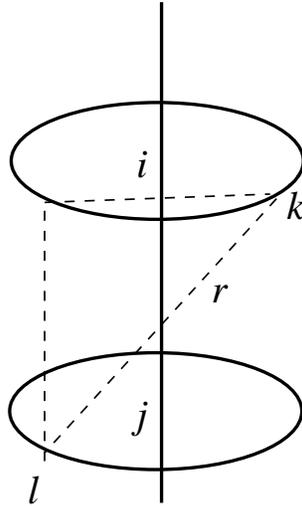,width=4cm,angle=0} }
\caption{ Distance between a surfactant located on the $i$'th ring
in the $k$'th position and the surfactant located on the $j$'th ring  and
the $l$'th position,
$r=b\sqrt{(i-j)^2+4R^2\sin^2\left(\pi
|k-l|/n\right)}$.}
\label{figure2}
\end{figure}

\item Hydrophobic interactions between the hydrocarbon tails:
\begin{equation}
\beta{\cal H}_3 = -\chi \sum_{i=1}^Z \sum_{j=1}^{n}
\sigma_{ij}\sigma_{i,j+1} -\chi \sum_{i=1}^{Z-1} \sum_j^n
\sigma_{ij}\sigma_{i+1,j} \;\;,
\label{e3} 
\end{equation}
\end{itemize}
where $\sigma_{i,n+1}=\sigma_{i1}$. The first term of
Eq.~(\ref{e3}) is due to the interactions between nearest
neighbor amphiphiles on the same ring, while the
second term is due the interactions between equivalent sites on
consecutive rings. The hydrophobicity parameter $\chi$ can be
related to the size of the amphiphiles alkyl chain~\cite{kuh98}.


The hydrophobic interactions between surfactant
and water produce an effective attraction between the amphiphiles, 
forcing them to stick together.  By doing so
they  expel the water molecules from their vicinity, lowering the
overall free energy.  The pairwise additive form of the hydrophobic
interaction adopted in Eq.~(\ref{e3}) is clearly an over simplification.
Nevertheless, we expect that this simple  expression will help
shed some light on the structure of the polyion-amphiphile complexes.


\section{Mean Field Theory}

To begin our study of the distribution of surfactants along
the polyion we shall appeal to the mean field theory~\cite{kuh98b,are00}.  
A note
of caution, however, must be raised.  While the mean field
theory is expected to work very well for Coulombic
long ranged interactions, it might not be so successful with
the short ranged hydrophobic forces.  This is in particular so
since the problem is intrinsically 
one dimensional, and the fluctuations associated
with the short ranged hydrophobic interactions are expected
to be significant~\cite{kuh99a}.  With this note of caution in mind 
we shall proceed with the mean field study.

The Gibbs-Bogoliubov inequality puts an upper bound on the total free
energy,  ${\cal F}\leq {\cal F}_0+\langle{\cal H}-
{\cal H}_0 \rangle_0\equiv {\cal \overline{F}}$, where ${\cal F}_0$ is
the free energy associated with the trial Hamiltonian ${\cal H}_0$.   
To perform the calculations
we shall take ${\cal H}_0$ to be of a particularly
simple one body form, 
\begin{equation}
{\cal H}_0=q\sum_i^Z \sum_p^n \phi_{ip} \sigma_{ip} \;.
\end{equation}
The partition function associated with ${\cal H}_0$ can
be calculated straight forwardly
\begin{equation}
Z_0=\sum_{\{\sigma\}} \exp\left\{ -\beta q\sum_{i,p} \phi_{ip}
\sigma_{ip}\right\} =\prod_{i,p} \left\{ 1+\exp\left(-\beta q \phi_{ip}
\right)\right\} \;\;,
\end{equation}
and the free energy is
\begin{equation}
{\cal F}_0=-\frac{1}{\beta}\sum_{i,p}\ln\left\{ 1+\exp\left(-\beta q \phi_{ip}
\right)\right\} \;\;.
\end{equation}
The average occupation of site $p$ on the ring $i$, 
$\rho_{ip}\equiv <\sigma_{ip}>_0$, is then
\begin{equation}
\rho_{ip}=\frac{1}{1+e^{\beta q\phi_{ip}}}
\end{equation}
or
\begin{equation}
\phi_{ip}=\frac{1}{\beta q}\ln\left(\frac{1-\rho_{ip}}{\rho_{ip}}\right)\;\;.
\end{equation} 
The free energy associated with  ${\cal H}_0$ can be rewritten
as
\begin{equation}
{\cal F}_0=
\frac{1}{\beta}\sum_{i,p} \ln(1-\rho_{ip}) \;\;.
\end{equation}

After evaluating the average of $\langle {\cal H}-{\cal H}_0\rangle_0$ 
with respect to ${\cal H}_0$, 
the upper bound to the total free energy becomes
\begin{eqnarray}
\beta {\cal \overline{F}} &=& 
\sum_{i,j}\ln( 1- \rho_{ij})
-\alpha \xi\sum_{i,j}^Z \sum_k^{n}
\frac{\rho_{ik}}{\sqrt{R^2+(i-j)^2}}
+ \frac{\alpha^2 \xi}{2}\sum_{i,j}^Z \sum_{k,l}^n
\frac{\rho_{ik}\rho_{jl}(1-\delta_{ij}\delta_{kl})}{
\sqrt{(i-j)^2+4R^2\sin^2\left(\frac{\pi}{n}
|k-l| \right)}} \nonumber \\
&& -\chi \sum_i^Z \sum_{j}^{n}
\rho_{ij}\rho_{i,j+1} -\chi \sum_{i=1}^{Z-1} \sum_j^n
\rho_{ij}\rho_{i+1,j}
-\sum_i^Z\sum_j^n\rho_{ij}\ln\left(\frac{1-\rho_{ij}}{
\rho_{ij}}\right) \;\;.
\label{free}
\end{eqnarray}

To find the optimum upper bound, Eq.~(\ref{free}) 
must be minimized with respect to the average  site occupation,
leading to
\begin{equation}
\rho_{ij}=\frac{1}{1+e^{\beta(\varphi_{ij}+\mu)}}\;,
\end{equation}
where $\mu$ is a Lagrange multiplier introduced to enforce the
constraint $\sum_{i,j}\rho_{ij}=m$ and
\begin{eqnarray}
\beta\varphi_{ij}&=&
-\alpha\xi \sum_k^Z\frac{1}{\sqrt{R^2+(i-k)^2}}
+ \alpha^2 \xi\sum_{k}^Z \sum_{l}^n
\frac{\rho_{kl}(1-\delta_{ik}\delta_{jl})}{
\sqrt{(i-k)^2+4R^2\sin^2\left(\frac{\pi}{n}
|j-l| \right)}} \nonumber \\
&& -\chi ( \rho_{i,j-1}+\rho_{i,j+1}) 
-\chi \left[ (1-\delta_{iZ})\rho_{i+1,j}+(1-\delta_{i1})\rho_{i-1,j}\right] \;.
\end{eqnarray}
Using the constraint above, $\mu$ can be evaluated leading to a
self consistent equation for the average site occupation,
\begin{equation}
\rho_{ij}=\frac{m}{m+\sum_{k,l}(1-\rho_{kl})e^{\beta(\varphi_{ij}-\varphi_{kl})}}\;\;.
\label{rho}
\end{equation}
Eq.~(\ref{rho}) can  be solved numerically to find the equilibrium
amphiphile distribution along the polyion.  

\begin{figure}
\begin{center}
\includegraphics[bb= 0 0 600 800, scale=0.3, angle= 270]{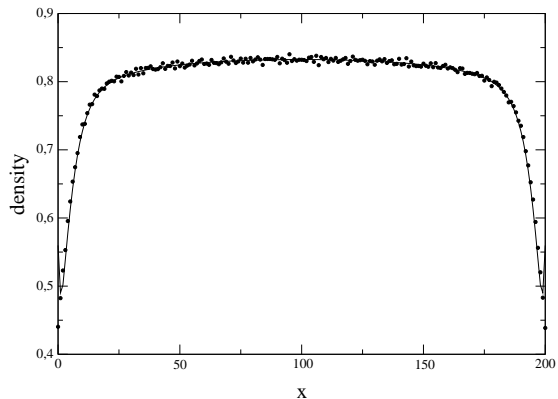}
\end{center}
\caption{Comparison between the profiles predicted by the mean field theory
and simulations results, 
for $\xi=4.17$, $\chi=1$, $Z=201$, $m=160$, and $n=5$.}
\label{fig:profilecomp}
\end{figure}

In Fig.~\ref{fig:profilecomp} we show the average number 
amphiphiles per ring along the polyion
for $\xi=4.17$ and $\chi=1$.  We note that the amphiphiles are
uniformly distributed along the polyion except at the ends of the
macromolecule, where their density is strongly depleted. 

\begin{figure}
\begin{center}
\includegraphics[bb= 0 0 600 800, scale=0.3, angle= 270]{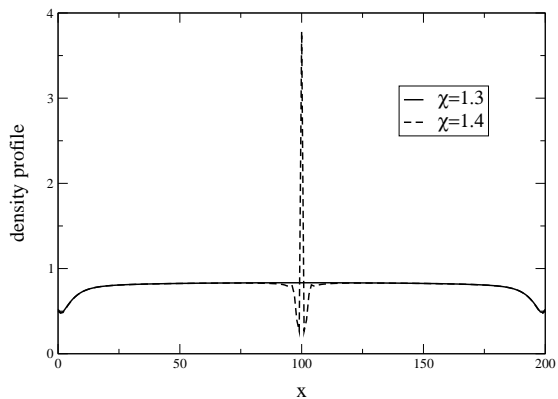}
\end{center}
\caption{The mean field theory prediction for $\chi=$1.3 and 1.4,
$\xi=4.17$, $Z=201$, $m=160$, $n=5$.    
The peak is  an artifact of the mean field approximation.}
\label{fig:peaks}
\end{figure}

For $\chi \approx 1.35$ a curious phenomenon occurs, as shown in
Fig.~\ref{fig:peaks}.  At this value
of hydrophobicity the sites of the central ring become preferentially 
occupied by the amphiphiles.  This corresponds 
to a micellization transition, in which the strong
hydrophobic attraction between the
surfactants overcomes entropy to produce a mesoscopic aggregate
of amphiphiles. As $\chi$ increases, a number of other
peaks appear. Within the mean field theory we find that
the micellization transition is of first order.  The fact that
a short ranged interaction produces a first order transition
in a pseudo one dimensional system should leave us
concerned.  To check the existence  of this transition we
have carried out a set of Monte-Carlo simulations (MC).    

\section{Simulations}

To simulate this model we use a standard Monte Carlo with particle-hole
exchange, not restricted to nearest-neighbors pairs. The density
profiles and the energy were measured after thermalization, results
being  both time and sample averaged. The simulations considered 
where done for DNA with $\xi=4.17$, $Z=201$, 
$n=160$, $m=5$ and various values of  $\chi$.  
The averages were obtained using  100  samples. 

\begin{figure}
\begin{center}
  \includegraphics[bb= 0 0 600 800, scale=0.3, angle=
  270]{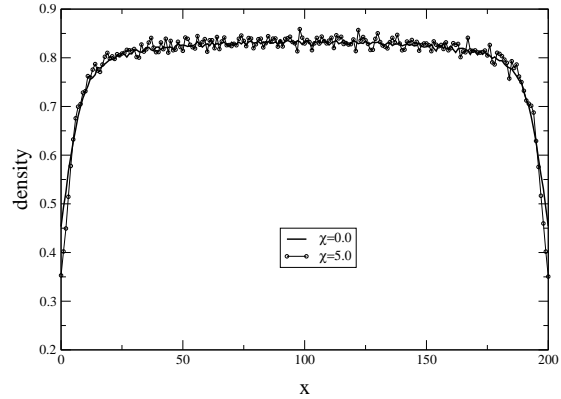}
\end{center}
\caption{The Monte Carlo density profiles for $\chi=0$ and $\chi=5$ 
for a system with
$\xi=4.17$, $Z=201$, $n=160$ and $m=5$. 
For $\chi=5$, the mean field
theory predicts the existence of many peaks, that are absent
in simulation.}
\label{fig:profilesmc}
\end{figure}

\begin{figure}
\begin{center}
  \includegraphics[bb= 0 0 600 800, scale=0.3, angle= 270]{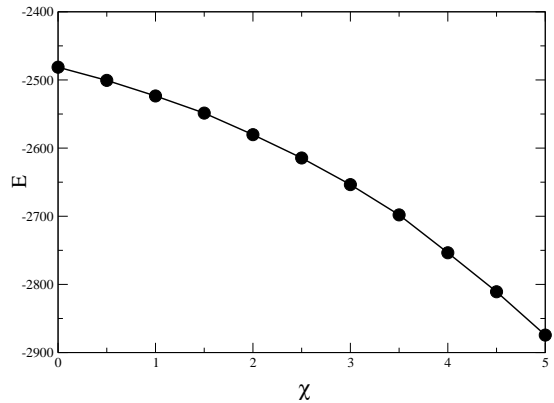}
\end{center}
\caption{The energy dependence on $\chi$ calculated using Monte Carlo,
$\xi=4.17$, $Z=201$, $n=160$ and $m=5$.}
\label{fig:energymc}
\end{figure}

For $\chi<1.35$ we find that the mean field theory is in excellent
agreement with the MC.  For $\chi>1.35$, on the other hand,
the simulations do not find any evidence of the  micellization transition
present in  mean field (see Fig.~\ref{fig:profilesmc}).  The energy
is a smooth function of $\chi$, with no indication of the first
order micellization transition, Fig.~\ref{fig:energymc}.  As expected,
the short ranged hydrophobic interaction can not
result in a phase transition in a one dimensional system. 

\begin{figure}
\begin{center}
  \includegraphics[bb= 0 0 600 800, scale=0.3, angle=
  270]{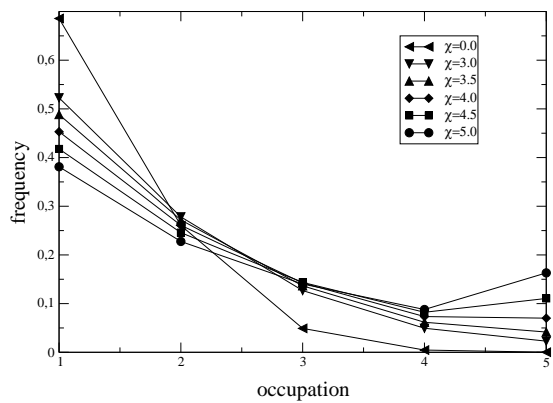}
\end{center}
\caption{Histogram  of cluster sizes for $Z$=201 and $n=160$, $\xi=4.17$.}
\label{fig:histo}
\end{figure}

In spite of the absence of a true phase transition, the correlations
between the condensed amphiphiles can lead to formation of
structures along the polyion.  To study these, we have
constructed a histogram of  amphiphile cluster sizes within 
the Monte Carlo simulation. Here the size of a cluster is
defined by the number of amphiphiles per ring. 
Fig.~\ref{fig:histo} shows  that for amphiphiles with short alkyl tails 
(small hydrophobicity) the clusters are composed 
of only one amphiphile,  
with larger aggregates being highly improbable.  With the increase in
$\chi$ we find, however, that this is no longer the case and a significant
fraction of amphiphiles belongs to the maximum sized cluster of 
$n$ amphiphiles.
Although this is not a thermodynamic transition, the change
in behavior evident in Fig.~\ref{fig:histo} can be associated with
the micellization.

\section{Conclusions}

We have studied a simple model of micellization in the presence
of polyelectrolyte.  It is found that the mean field theory
predicts a first order micellization transition for the ionic
amphiphiles condensed on a polyion.  This thermodynamic 
transition is an artifact of the mean field approximation
and is the result of neglect of fluctuations associated
with the short ranged hydrophobic interactions.  Monte Carlo
simulations show that the mean field works very well for small
hydrophobicities, but fails completely for strong short ranged 
interactions.  Indeed the simulations do not find any evidence
of a phase transition.  Nevertheless, 
if the hydrophobic interaction are sufficiently strong they will
lead to significant correlations between
the condensed amphiphiles, which can be interpreted as
a micellar formation along the polyion chain. 

\bigskip

{\bf ACKNOWLEDGMENTS}

\bigskip
This work was 
supported in part by CNPq and Fapergs, Brazilian science agencies.


\end{document}